\documentclass[12pt]{article}
\usepackage[cp1251]{inputenc}
\usepackage{amsmath}
\usepackage{amsfonts}
\usepackage{amssymb}
\def\pa{\partial}                       
\def\beq{\begin{eqnarray}}    
\def\eeq{\end{eqnarray}}      

\def\vep{\varepsilon}

\def\al{\alpha}

\def\be{\begin{equation}}
\def\ee{\end{equation}}

\def\bea{\begin{eqnarray}}
\def\eea{\end{eqnarray}}

\def\Box{\square}
\begin{document}
\date{}

\begin{center}
{\Large\textbf{On gauge-invariant deformation of reducible
gauge theories}}

\vspace{18mm}

{\large P.M. Lavrov$^{(a,b)} \footnote{E-mail:
lavrov@tspu.edu.ru}$,\;
}

\vspace{8mm}

\noindent  ${{}^{(a)}}
${\em
Center of Theoretical Physics, \\
Tomsk State Pedagogical University,\\
Kievskaya St.\ 60, 634061 Tomsk, Russia}

\noindent  ${{}^{(b)}}
${\em
National Research Tomsk State  University,\\
Lenin Av.\ 36, 634050 Tomsk, Russia}

\vspace{20mm}

\begin{abstract}
\noindent
New method for construction of gauge-invariant deformed theory
from an initial gauge theory proposed in our previous papers \cite{BL-1,BL-2} for
closed/open gauge algebras is extended to the case of reducible gauge algebras.
The deformation procedure is explicitly described with
the help of generating functions of anticanonical transformations depending on fields of
the initial gauge action only. The deformed gauge-invariant action and the deformed
gauge generators are described with the help of the generating functions
in a closed and simple
form.
As an example of reducible gauge systems we consider the free fermionic p-form fields
or, in another words, the antisymmetric tensor-spinor fields.
It is proved that gauge-invariant deformation of fermionic p-form fields leads
always to non-local deformed theory which does not contain a closed local sector.
In its turn the model based on two fermionic 2-form fields and a real massive
scalar field admits
local interactions between these fields in local sector of the deformed action.

\end{abstract}

\end{center}

\vfill

\\

\newpage

\section{Introduction}
\noindent
Recently, a new approach to  gauge-invariant deformation of gauge theories has been proposed
in our papers \cite{BL-1}, \cite{BL-2}. This approach is closely related with
the Batalin-Vilkovisky (BV) formalism \cite{BV}, \cite{BV1}, \cite{BV2} which is the most powerful method for
covariant quantization of general gauge theories.
The central role in the BV formalism belongs to  the classical master equation
formulated in terms of the antibracket. It is a remarkable fact that
the antibracket is invariant under anticanonical transformations that  helps
in studying different properties of gauge theories \cite{VLT}, \cite{BLT-15}, \cite{BL-16},
\cite{BLT-21}, \cite{AT}. In this connection, it seems useful
to remind the standard approach to the
problem of gauge-invariant deformation for systems with gauge invariance.

Construction of consistent interactions among fields with a gauge freedom or
  gauge-invariant deformations of a free gauge system  is formulated as follows \cite{BH}.
Starting point of deformation procedure is  a given  theory
described by an action $S_0=S_0[A]$ of field $A=\{A^i\}$ which is supposed to be
invariant under gauge transformations,
\beq
S_{0,i}R^i_{0\alpha}=0,\quad \delta A^i=R^i_{0\alpha}\xi^{\alpha},
\eeq
where $\xi^{\alpha}$ are arbitrary functions of space-time coordinates.
It is required to construct a final (deformed) action $S$ as
\beq
S=S_0+gS_1+g^2S_2+\cdots,
\eeq
where $g$ is a deformation parameter,  in such a way
that initial gauge generators $R^i_{0\alpha}$
are deformed,
\beq
R^i_{0\alpha}\quad \rightarrow \quad R^i_{\alpha}=R^i_{0\alpha}+
gR^i_{1\alpha}+g^2R^i_{2\alpha}+\cdots ,
\eeq
to final gauge generators $R^i_{\alpha}$  \cite{BH},\cite{H} so that the
deformed action $S$ is invariant under the deformed gauge symmetry,
\beq
S_{,i}R^i_{\alpha}=0 .
\eeq
To arrive these results it has been proposed \cite{BH},\cite{H} to embed  the deformation
procedure in the BV formalism as a part of solutions
to the classical master equation for an action ${\cal S}$,
\beq
({\cal S},{\cal S})=0 ,
\eeq
with the boundary condition
\beq
 {\cal S}\big|_{g= 0}=S_0.
\eeq
The bridge connecting solutions ${\cal S}$  to the classical
master equation with the deformed action $S$ is established
with the help of the relation
\beq
 {\cal S}\big|_{{\rm antifields} \ = \ 0}=S.
\eeq
Solutions to the classical master equation are searched in
the form of Taylor expansion with respect to parameter $g$,
\beq
 {\cal S}=S_0+g{\cal S}_1+g^2{\cal S}_2+\cdots .
\eeq
Then, the classical master equation for action  ${\cal S}$ generates
the infinite set of equations
\beq
(S_0,S_0)=0, \quad (S_0,{\cal S}_1)=0,\quad
2(S_0,{\cal S}_2)+({\cal S}_1,{\cal S}_1)=0, \quad \cdots .
\eeq
Usually, this system of equations is analyzed with the help of the cohomological approach
\cite{BH},\cite{H} (see also recent applications \cite{D}, \cite{SS}).
In general, this approach to the deformation procedure does not give
a possibility to present the deformed action and the deformed gauge generators
in an explicit and closed form. In fact, it was a reason for us to reconsider the
gauge-invariant deformations of gauge systems within the BV formalism using
the  invariance of antibracket under anticanonical transformations
\cite{BL-1}, \cite{BL-2}.

The anticanonical transformations by itself can be described in two ways, namely,
in terms of generating functionals or with the help of generators.
Being equivalent on theoretical level, they might be distinguished
in practical applications. It happened really in our reformulation
of the deformation procedure. The description of anticanonical transformations
with the help of generating functionals is more preferred. Moreover,
it was realized \cite{BL-2}
that the deformation problem can be solved using the so-called minimal
anticanonical transformations in the minimal antisymplectic space
when the corresponding functionals are described with the help
of generating functions depending on fields of initial theory
in the number equals to the number of initial fields
and having the same transformation properties as initial fields.
The gauge-invariant deformations in papers \cite{BL-1}, \cite{BL-2}
have been solved for initial gauge theories with closed/open algebras.
Main goal of present paper is to extend the new approach for reducible
gauge theories.

The paper is organized as follows. In section 2, we  review
the basic notions of reducible gauge theories and corresponding gauge algebras
underlying such gauge systems.
Section 3 is devoted to presentation  of the deformed gauge action and
corresponding deformed
gauge symmetry  in terms of a single
generating function depending on initial fields only.
In section 4, we consider the free fermionic p-form  fields as an example
of reducible gauge system subjected to gauge-invariant deformations.
In section 5, local interactions of fermionic 2-form fields
and a real massive scalar field as
the result of suitable deformation of the initial free model of these fields
are constructed.
In section 6, we summarize the results.

In the paper, we systematically use the DeWitt's condensed notations \cite{DeWitt}
and employ the symbols $\varepsilon(A)$ for the Grassmann parity and
${\rm gh}(A)$ for the ghost number, respectively. The right and left
functional derivatives are marked by special symbols $"\leftarrow"$
and $"\overrightarrow{}"$ respectively. Arguments of any functional
are enclosed in square brackets $[\;]$, and arguments of any
function are enclosed in parentheses, $(\;)$. The symbol $F_{,i}(A)$ is used
for right partial derivative of function $F(A)$ with respect to $A^i$.

\section{Reducible gauge theories}
\noindent
We consider a gauge theory of the fields $A=\{A^i\}$ with Grassmann
parities $\varepsilon(A^i)=\varepsilon_i$ and ghost numbers $\;{\rm
gh}(A^i)=0$. The theory is described by the initial action $S_0[A]$
and gauge generators $R^i_{\alpha}(A)$
($\varepsilon(R^i_{\alpha}(A))=\varepsilon_i+\varepsilon_{\alpha},\;
{\rm gh}(R^i_{\alpha}(A))=0$). The action is invariant under the
gauge transformations
\beq
\label{ggen0}
\delta A^i=R^i_{\alpha}(A)\xi^{\alpha},
\eeq
where the gauge parameters
$\xi^{\alpha}$ ($\varepsilon(\xi^{\alpha})=\varepsilon_{\alpha}$)
are the arbitrary functions of space-time coordinates. Condition of
gauge invariance is written in the standard form \footnote{To simplify presentation of all
relations containing the right functional derivative of functional $S_0[A]$ with respect
to field $A^i$ we will use the symbol $S_{0,i}[A]=S_{0,i}$ .}
\beq
\label{S0giv}
S_{0,i}[A]R^i_{\alpha}(A)=0, \quad \alpha = 1,2,..., m.
\eeq
It is assumed that the fields $A=\{A^i\}$ are linear independent with
respect to the index $i$ however, in general, these generators may
be linear dependent with respect to index $\alpha$.
Linear dependence of $R^i_{\alpha}(A)$ implies that the matrix
 $R^i_{\alpha}(A)$ has at the extremals $S_{0,j}[A]=0$ zero-eigenvalue
 eigenvectors $Z^{\alpha}_{\alpha_1}=Z^{\alpha}_{\alpha_1}(A)$, such
 that
\begin{eqnarray}
\label{GTR1}
 R^i_{\alpha}(A)Z^{\alpha}_{\alpha_1}(A) =S_{0,j}[A]K^{ji}_{\alpha_1}(A),\quad
 \alpha_1 = 1, ..., m_1,
\end{eqnarray}
 and the number $\varepsilon_{\alpha_1}=0,1$ can be found in such a way
 that $\varepsilon(Z^{\alpha}_{\alpha_1})=\varepsilon_{\alpha}+
 \varepsilon_{\alpha_1}$. Matrices $K^{ij}_{\alpha_1}=K^{ij}_{\alpha_1}(A)$
 in (\ref{GTR1})
 can be chosen to possess the properties:
\begin{eqnarray}
\nonumber
 K^{ij}_{\alpha_1}=-(-1)^{\varepsilon_i\varepsilon_j}K^{ji}_{\alpha_1}
 ,\quad
\varepsilon(K^{j\;\!i}_{\alpha_1}) =
\varepsilon_i+\varepsilon_j+\varepsilon_{\alpha_1}.
\end{eqnarray}

The generators $R^i_{\alpha}(A)$ satisfy the
following relations
\beq
\label{ga} R^i_{\alpha ,
j}(A)R^j_{\beta}(A)-(-1)^{\varepsilon_{\alpha}\varepsilon_
{\beta}}R^i_{\beta ,j}(A)R^j_{\alpha}(A)=-R^i_{\gamma}(A)
F^{\gamma}_{\alpha\beta}(A) - S_{0,j}[A]M^{ji}_{\alpha\beta}(A),
\eeq
where
$F^{\gamma}_{\alpha\beta}(A)=F^{\gamma}_{\alpha\beta}$
($\varepsilon(F^{\gamma}_{\alpha\beta})=\varepsilon_{\alpha}+
\varepsilon_{\beta}+\varepsilon_{\gamma},\;{\rm
gh}(F^{\gamma}_{\alpha\beta})=0$) are the structure coefficients
depending, in general, on the fields $A^i$ with the following
symmetry properties $F^{\gamma}_{\alpha\beta}=
-(-1)^{{\varepsilon_{\alpha}\varepsilon_{\beta}}}F^{\gamma}_{\beta\alpha}$,
and $M^{ij}_{\alpha\beta}(A)=M^{ij}_{\alpha\beta}$ satisfy the conditions
\beq
M^{ij}_{\alpha\beta} = -(-1)^{\varepsilon_i\varepsilon_j}
M^{ji}_{\alpha\beta} =
-(-1)^{\varepsilon_{\alpha}\varepsilon_{\beta}}M^{ij}_{\beta\alpha}.
\eeq

In its turn, the set $Z^{\alpha}_{\alpha_1}$ may be
linearly dependent as itself, so that at the
 extremals $S_{0,i}=0$ there exists the set of zero-eigenvalue eigenvectors
 $Z^{\alpha_1}_{\alpha_2}=Z^{\alpha_1}_{\alpha_2}(A)$
\begin{eqnarray}
\label{GTR2}
 Z^{\alpha}_{\alpha_1}Z^{\alpha_1}_{\alpha_2} =
 S_{0,j}L^{j\alpha}_{\alpha_2},\quad
 \alpha_2 = 1, ..., m_2
\end{eqnarray}
 and numbers $\varepsilon_{\alpha_2}=0,1$ such that
 $\varepsilon(Z^{\alpha_1}_{\alpha_2})=\varepsilon_{\alpha_1} +
\varepsilon_{\alpha_2}$.
In the general case the set
$Z^{\alpha_1}_{\alpha_2}$ can be redundant and so on.
In such a way the sequence of reducibility equations arises:
\begin{eqnarray}
\label{GTRS}
 Z^{\alpha_{s-2}}_{\alpha_{s-1}}Z^{\alpha_{s-1}}_{\alpha_s} =
 S_{0,j}L^{j\alpha_{s-2}}_{\alpha_s},\quad
 \alpha_s = 1, ..., m_s; s=1,.., L,
\end{eqnarray}
 where the following notations are introduced:
\begin{eqnarray}
Z^{\alpha_1}_{\alpha_0}\equiv R^i_{\alpha},\quad
 L^{j\alpha_1}_{\alpha_0}\equiv K^{j\;\!i}_{\alpha},\quad
\varepsilon(Z^{\alpha_{s-1}}_{\alpha_s}) = \varepsilon_{\alpha_{s-1}}+
\varepsilon_{\alpha_{\alpha_s}}.
\end{eqnarray}
If the set $\{Z^{\alpha_{L-1}}_{\alpha_L}\}$ is linear independent then
one meets with a gauge theory of $L$-stage reducibility.

The set of gauge generators $\{R^i_{\alpha}\}$,
eigenvectors $\{Z^{\alpha_{s-1}}_{\alpha_s}\}$
and structure functions
$\{L^{j\alpha_{s-2}}_{\alpha_s}\}$  defines the
structure of gauge algebra on the first level.
For irreducible theories, the structure of gauge algebra on the second
level is defined  by the set of structure functions
$\{F^{\gamma}_{\alpha\beta}\}$ and matrices $\{M^{ij}_{\alpha\beta}\}$
in Eq.~(\ref{ga}). For reducible theories, the existence of relations
among the $Z^{\alpha_{s-1}}_{\alpha_s}$ (\ref{GTRS}) leads to the
appearance of new structure functions. Let us demonstrate this point
for a first-stage reducible gauge theory. To this end, let us multiply
the relation (\ref{ga}) by the eigenvector $Z^{\beta}_{\alpha_1}$.
We obtain
\begin{eqnarray}
\label{GAR}
 \bigg(R^i_{\alpha , j}R^j_{\beta}-
 (-1)^{\varepsilon_{\alpha}\varepsilon_
{\beta}}R^i_{\beta ,j}R^j_{\alpha}+R^i_{\gamma}
F^{\gamma}_{\alpha\beta}+ S_{0,j}M^{j\;\!i}_{\alpha\beta}\bigg)
Z^{\beta}_{\alpha_1}=0.
\end{eqnarray}
First, note that relations (\ref{GTR1}) allows us to express
$R^j_{\beta}Z^{\beta}_{\alpha_1}$ as a term proportional to the
equations of motion. Second, by differentiating  Eqs.~(\ref{GTR1})
and (\ref{S0giv})
with respect to $A$ one obtains that
\begin{eqnarray}
\label{GAR1}
R^i_{\beta , j}Z^{\beta}_{\alpha_1}
(-1)^{\varepsilon_j(\varepsilon_{\beta}+\varepsilon_{\alpha_1})}
+R^i_{\beta}Z^{\beta}_{\alpha_1,j}=S_{0,jl}K^{li}_{\alpha_1}
 (-1)^{\varepsilon_j(\varepsilon_i +\varepsilon_{\alpha_1})}+
 S_{0,l}K^{li}_{\alpha_1,j},
\end{eqnarray}
\begin{eqnarray}
\label{GAR2}
S_{0,ji}R^j_{\alpha}(-1)^{\varepsilon_l\varepsilon_{\alpha}}+
S_{0,i}R^i_{\alpha, j}=0.
\end{eqnarray}
Then, multiplying Eqs.~(\ref{GAR1}) by $R^j_{\alpha}$, using the
Noether identities (\ref{S0giv}) and relations (\ref{GAR2}),
we find
\begin{eqnarray}
\label{}
\nonumber
 - (-1)^{\varepsilon_{\alpha}\varepsilon_{\beta}}
  R^i_{\beta ,j}R^j_{\alpha}Z^{\beta}_{\alpha_1}=
  (-1)^{\varepsilon_{\alpha}\varepsilon_{\alpha_1}}
 R^i_{\beta}Z^{\beta}_{\alpha_1,j}R^j_{\alpha}+
 S_{0,j}(R^j_{\alpha ,l}K^{il}_{\alpha_1}
 (-1)^{\varepsilon_{\alpha}\varepsilon_i} -
 K^{ij}_{\alpha_1,l}R^l_{\alpha}
 (-1)^{\varepsilon_{\alpha}\varepsilon_{\alpha_1}}).
\end{eqnarray}
Returning with this result into (\ref{GAR}), one can obtained the
relations
\begin{eqnarray}
\label{}
\nonumber
 R^i_{\beta}\big((-1)^{\varepsilon_{\alpha}\varepsilon_{\alpha_1}}
 Z^{\beta}_{\alpha_1,j}R^j_{\alpha}-
 F^{\beta}_{\alpha\gamma}Z^{\gamma}_{\alpha_1}\big)=
S_{0,j}Y^{ji}_{\alpha_1\alpha}
\end{eqnarray}
where all terms proportional to the equation of motion have been
collected into $Y^{ji}_{\alpha\alpha_1}$.
Taking into account the completeness of the set of eigenvectors
$Z^{\alpha}_{\alpha_1}$,  the general solution to this equation,
\begin{eqnarray}
\label{GAnF2}
 (-1)^{\varepsilon_{\alpha}\varepsilon_{\alpha_1}}
 Z^{\beta}_{\alpha_1,j}R^j_{\alpha}-
 F^{\beta}_{\alpha\gamma}Z^{\gamma}_{\alpha_1}=
 - Z^{\beta}_{\beta_1}P^{\beta_1}_{\alpha_1\alpha}-
S_{0,j}Q^{j\beta }_{\alpha_1\alpha},
\end{eqnarray}
defines a new gauge-structure relation similar to Eq.~(\ref{ga}).
Therefore, two new structure functions $P^{\beta_1}_{\alpha\alpha_1}$
and $Q^{j\beta }_{\alpha\alpha_1}$ arise to complete definition of the
structure of gauge algebra for the first-stage reducible theory on
the second level. To define the structure of gauge algebra on the third level, one has
to consider the Jacobi identity for gauge transformations and some
consequences from gauge-structure relations of previous levels.

In principal, there is no problem in deriving the corresponding gauge algebra for
reducible gauge theories of any stage of reducibility but here we omit further calculations.
Let us remark only that, in general, the structure of gauge algebra looks
like a set of infinite
number of structure functions which define infinite number of gauge-structure relations.
It is remarkable
fact that all these relations can be collected within the BV method
in a solution to the classical master equation.

Within the BV formalism, studies of classical aspects of gauge-invariant deformations
can be performed in the
minimal antisymplectic space of fields $\phi^A$ and antifields $\phi^*_A$
as it was pointed out in \cite{BL-2}.
For reducible $L$-stage gauge theory
of fields $A^i$,
it  contains main chains of the ghost $C^{\al_s}_s$,
and pyramids of the ghost for ghost $C^{\al_s}_{s(n_s)}$,
\begin{eqnarray}
\label{ConfSpaceRBV}
\phi^A = \left(A^i;\; C^{\al_s}_s, s=0,1,..., L,
C^{\al_s}_{s(n_s)}, s=1,...,L, n_s=1,...,s\right)
\end{eqnarray}
with the properties
\begin{eqnarray}
\label{}
\nonumber
\varepsilon (C^{\al_s}_s) & =&
(\varepsilon_{\al_s}+s+1)\;{\rm mod 2}, \quad s=0,1,...,L,\\
\nonumber
\vep(C^{\al_s}_{s(n_s)})&=&(\vep_{\al_s}+s +1)\;{\rm mod 2},
\quad s=1,...,L,\; \quad n_s=1,...,s,
\\
\nonumber
gh(C^{\al_s}_s)&=&(s+1), \; s=0,1,...,L\\
gh(C^{\al_s}_{s(n_s)})&=& s+1-2n_s,\; s=1,...,L,
\;n_s=1,...,s ,
\end{eqnarray}
and the corresponding set of antifields
\begin{eqnarray}
\label{antiSpaceRBV} \phi^*_A &=& \Big(A^*_i,
C^*_{s{\al_s}}, s=0,1,..., L,
\;\;\;\;C^*_{s(n_s)\al_s}, s=1,...,L, n_s=1,...,s\Big) .
\end{eqnarray}
The statistics of $\phi^*_A$ is opposite to the statistics of the
corresponding fields $\phi^A$
\begin{eqnarray}
\varepsilon (\phi^*_A) = \varepsilon_A + 1 ,
\nonumber
\end{eqnarray}
and the ghost numbers of fields and corresponding antifields are connected
by the rule
\bea
\nonumber
\quad gh(\phi^*_A)=-1 -gh(\phi^A).
\eea

In comparison with original proposal of Ref.~\cite{BV2}, we have slightly
(for simplicity and uniformity) changed notation of pyramids of fields.
As an example, for a second-stage reducible theory, the following
identification for the pyramids of fields exists:
\bea
\nonumber
C^{'\al_1}_1&\equiv& C^{\al_1}_{1(1)},\quad C^{'\al_2}_2\equiv
C^{\al_2}_{2(1)}.
\eea

The basic object of the BV formalism is the extended action
$S=S[\phi,\phi^*]$ satisfying the classical master equation,
\beq
\label{master} (S,S)=0,
\eeq
and the boundary condition,
\beq
\label{boundary} S[\phi,\phi^*]\Big|_{\phi^*=0}=S_0[A].
\eeq
The classical master equation (\ref{master}) is written in terms of antibracket
which is defined for any functionals $F[\phi,\phi^*]$ and
$H[\phi,\phi^*]$ in the form
\beq
\label{antibracket} (G,H)=
G\left(\overleftarrow{\pa}_{\!\!\phi^A}\overrightarrow{\pa}_{\!\!\phi^*_{A}}-
\overleftarrow{\pa}_{\!\!\phi^*_{A}}\overrightarrow{\pa}_{\!\!\phi^A}\right)H.
\eeq

The gauge invariance of the initial action $S_0[A]$ leads to
invariance of the action $S[\phi,\phi^*]$,
\beq
\delta_B S=0 ,
\eeq
under the global supersymmetry transformations (BRST transformations
\cite{brs1}, \cite{t})
\beq
\delta_B \phi^A=(\phi^A,S)\mu=
\overrightarrow{\pa}_{\!\!\phi^*_{A}}S\;\mu,\quad \delta_B
\phi^*_{A}=0,
\eeq
as a consequence that $S$ satisfies the classical
master equation. Here, $\mu$ is a constant Grassmann parameter.
We emphasize that the antibracket is a key element of compact description of the
classical gauge theories within the BV formalism. An important
property of the antibracket (\ref{antibracket}),
is its invariance with respect to anticanonical transformations of
fields and anti-fields \cite{BV,BV1}. It leads to statement that any
two solutions of classical master equation (\ref{master}) are related
one to another by some anticanonical transformation.

\section{Deformed action}

New approach to gauge-invariant deformation of a gauge theory was proposed in our papers
\cite{BL-1}, \cite{BL-2} for theories with the closed/open gauge algebras.
Here,  we are going to generalize the results for theories
when the gauge algebra is reducible.

Classical aspects of the gauge-invariant deformation
of initial theory can be studied in the minimal antisymplectic space using the minimal
anticanonical transformations as it was proved in \cite{BL-2}.
It means that the  anticanonical transformations
\beq
\label{transformation}
\phi^*_{A}=Y[\phi,\Phi^*]\overleftarrow{\pa}_{\!\!\phi^A},\quad
\Phi^A=\overrightarrow{\pa}_{\!\!\Phi^*_A}Y[\phi,\Phi^*],
\eeq
where
$Y=Y[\phi,\Phi^*]$ ($\varepsilon(Y)=1,\;{\rm gh}(Y)=-1$) is the
generating functional
 are non-trivial in the sector
of minimal antisymplectic space only
\beq
\label{antiCT}
Y[\phi,\Phi^*]=\Phi^*_A\phi^A+ {\cal A}^*_ih^i(A).
\eeq
Here, $h^i(A)=h^i$
$(\varepsilon(h^i)=\varepsilon_i,\; {\rm gh}(h^i)=0)$
are  arbitrary functions of
fields $A^i$ having the same transformation laws as for $A^i$.

For simplicity of presentation and notations without loss of generality
of all conclusions and statements,   we restrict ourselves to the case of
first-stage reducibility of the initial action when in
(\ref{transformation}) and (\ref{antiCT}) $\phi^A=(A^i,C^{\alpha},C^{\alpha_1}),\;
\phi^*_{A}=(A^*_i,C^*_{\alpha}, C^*_{\alpha_1})$ and
$\Phi^A=({\cal A}^i,{\cal C}^{\alpha},{\cal C}^{\alpha_1}),\;
\Phi^*_{A}=({\cal A}^*_i,{\cal C}^*_{\alpha}, {\cal C}^*_{\alpha_1})$.
Taking into account the gauge invariance of the initial action
(\ref{ggen0}) and the boundary condition (\ref{boundary}), one can
write the action $S=S[\phi,\phi^*]$ up to the terms linear in
antifields in the form
\beq
\label{S}
S=S_0[A]+A^*_iR^i_{\alpha}(A)C^{\alpha}+
C^*_{\gamma}\Big(Z^{\gamma}_{\alpha_1}(A)C^{\alpha_1}_1-
\frac{1}{2}F^{\gamma}_{\alpha\beta}(A)C^{\beta}C^{\alpha}
(-1)^{\varepsilon_{\alpha}}\Big)+O(\phi^{*\;2}).
\eeq

Making use of the anticanonical transformations (\ref{antiCT}) in the action
(\ref{S}), we obtain  the functional
$\widetilde{S}=\widetilde{S}[\phi,\phi^*]=S[\Phi(\phi, \phi^*), \Phi^*(\phi, \phi^*)]$
which satisfies the
classical master equation
\beq \label{dactS}
(\widetilde{S},\widetilde{S})=0 ,
\eeq
and has the following form up to the terms linear in antifields
\beq \label{dfact}
\widetilde{S}=\widetilde{S}_0[A]+A^*_i\widetilde{R}^i_{\alpha}(A)C^{\alpha}+
C^*_{\gamma}\Big(\widetilde{Z}^{\gamma}_{\alpha_1}(A)C^{\alpha_1}_1-
\frac{1}{2}\widetilde{F}^{\gamma}_{\alpha\beta}(A)C^{\beta}C^{\alpha}
(-1)^{\varepsilon_{\alpha}}\Big)+O(\phi^{*\;2}),
\eeq
where the quantities
\beq\nonumber
&&\widetilde{S}_0[A]=S_0[A+h(A)],\quad
\widetilde{R}^i_{\alpha}(A)=(M^{-1}(A))^i_{\;j}R^j_{\alpha}(A+h(A)), \\
\label{dq}
&&
\widetilde{F}^{\gamma}_{\alpha\beta}(A)=F^{\gamma}_{\alpha\beta}(A+h(A)),\quad
\widetilde{Z}^{\alpha}_{\alpha_1}(A)=Z^{\alpha}_{\alpha_1}(A+h(A))
\eeq
present the deformed initial action, $\widetilde{S}_0[A]$,
the deformed gauge generators, $\widetilde{R}^i_{\alpha}(A)$,
the deformed structure coefficients, $\widetilde{F}^{\gamma}_{\alpha\beta}(A)$,
 and
the deformed eigenvectors, $\widetilde{Z}^{\alpha}_{\alpha_1}(A)$.
The matrix $(M^{-1}(A))^i_{\;j}$
is inverse to
\beq
\label{M}
M^i_{\;j}(A)=\delta^i_{\;j}+h^i_{\;,j}(A) ,
\eeq
The action $\widetilde{S}_0[A]$ is invariant under the gauge transformations
$\delta A^i=\widetilde{R}^i_{\alpha}(A)\xi^{\alpha}$,
\beq
\widetilde{S}_{0,i}[A]\widetilde{R}^i_{\alpha}(A)=0.
\eeq
Therefore, the main problem of gauge-invariant deformation of a given gauge system
has the explicit closed solution as for the deformed action as well as for
deformed gauge generators. Such solutions  is described in terms of generating function
$h(A)$ only. The first relations in deformed gauge algebra for first-stage reducible theories
read
\beq
\nonumber
&&\widetilde{R}^i_{\alpha ,
j}(A)\widetilde{R}^j_{\beta}(A)-(-1)^{\varepsilon_{\alpha}\varepsilon_
{\beta}}\widetilde{R}^i_{\beta ,j}(A)\widetilde{R}^j_{\alpha}(A)=
-\widetilde{R}^i_{\gamma}(A)
\widetilde{F}^{\gamma}_{\alpha\beta}(A)-\widetilde{S}_{0,j}[A]
\widetilde{M}^{ji}_{\alpha\beta}(A),\\
\label{dga}
&&\widetilde{R}^i_{\alpha}(A)\widetilde{Z}^{\alpha}_{\alpha_1}(A)=
\widetilde{S}_{0, j}[A]\widetilde{K}^{ji}_{\alpha_1}(A),
\eeq
where the functions $\widetilde{M}^{ji}_{\alpha\beta}(A)$ and
$\widetilde{K}^{ji}_{\alpha_1}(A)$ are
\beq
&& \widetilde{M}^{ji}_{\alpha\beta}(A)=-
(M^{-1}(A))^j_{\;l}(M^{-1}(A))^i_{\;k}M^{kl}_{\alpha\beta}(A+h(A))
(-1)^{\varepsilon_l\varepsilon_i}, \\
&&\widetilde{K}^{ji}_{\alpha_1}(A)=-
(M^{-1}(A))^j_{\;l}(M^{-1}(A))^i_{\;k}K^{kl}_{\alpha_1}(A+h(A))
(-1)^{\varepsilon_l\varepsilon_i},
\eeq
In the same manner, we deduce the relations which define
the structure functions of deformed algebra on the second level
\beq
\label{GAnF2d}
 (-1)^{\varepsilon_{\alpha}\varepsilon_{\alpha_1}}
 \widetilde{Z}^{\beta}_{\alpha_1,j}(A)\widetilde{R}^j_{\alpha}(A)-
 \widetilde{F}^{\beta}_{\alpha\gamma}(A)\widetilde{Z}^{\gamma}_{\alpha_1}(A)=
 - \widetilde{Z}^{\beta}_{\beta_1}(A)\widetilde{P}^{\beta_1}_{\alpha_1\alpha}(A)-
\widetilde{S}_{0,j}[A]\widetilde{Q}^{j\beta }_{\alpha_1\alpha}(A),
\eeq
where
\beq
\widetilde{P}^{\beta_1}_{\alpha_1\alpha}(A)=
P^{\beta_1}_{\alpha_1\alpha}(A+h(A)),\qquad
\widetilde{Q}^{j\beta }_{\alpha_1\alpha}(A)=
(M^{-1}(A))^j_{\;k}Q^{k\beta }_{\alpha_1\alpha}(A+h(A)).
\eeq

The action (\ref{dfact}) is invariant under the BRST transformations,
\beq
\delta_B \widetilde{S}=0,\quad
\delta_B \phi^A=(\phi^A,\widetilde{S})\mu=
\overrightarrow{\pa}_{\!\!\phi^*_{A}}\widetilde{S}\;\mu,\quad
\delta_B \phi^*_{A}=0.
\eeq
and, therefore, the deformed theory repeats all basic properties of the initial
system on quantum level.

From the analysis carried out, the following conclusions can be drawn:
1) for any reducible theory with an gauge-invariant action $S_0[A]$
the deformed gauge-invariant action, $\widetilde{S}_0[A]$, is described by the
formula $\widetilde{S}_0[A]=S_0[A+h(A)]$ where $h(A)$ is a generating
function of the anticanonical transformation, 2) the deformed gauge generators,
$\widetilde{R}^i_{\alpha}(A)$, are defined
through the initial ones, $R^i_{\alpha}(A)$, by the relations (\ref{dq}),
3) the chain of deformed eigenvectors,
$\widetilde{Z}^{\alpha_{s-1}}_{\alpha_s}(A), s=1,2,...,L$, is expressed
in the form $\widetilde{Z}^{\alpha_{s-1}}_{\alpha_s}(A)=
Z^{\alpha_{s-1}}_{\alpha_s}(A+h(A))$, 4) the structure quantities appearing in
relations (\ref{GTRS}) are deformed by the rule
$\widetilde{L}^{j\alpha_{s-2}}_{\alpha_s}(A)=(M^{-1}(A))^j_{\;k}
L^{k\alpha_{s-2}}_{\alpha_s}(A+h(A))$,
5) the deformed gauge algebra looks like as initial
gauge algebra in which all structure coefficients are replaced by the deformed ones, 6)
the same conclusion is valid for relation between actions $S$ and $\widetilde{S}$  satisfying
the classical master equation.

\section{On deformation of fermionic p-form fields}
\noindent
As an example of reducible theories,
we  consider antisymmetric tensor-spinor fields or, in another words,
fermionic p-form fields, $\psi^a_{µ_1µ_2...µ_p}$,
where $a$ is a spinor index and the $µ_i$ are space-time
indices. The fields $\psi^a_{µ_1µ_2...µ_p}$
 are totally antisymmetric in their space-time indices:
\beq
\psi^a_{µ_1µ_2...µ_p}=\psi^a_{[µ_1µ_2...µ_p]}.
\eeq
Anti-symmetrization of tensor $A_{µ_1µ_2...µ_p}$ is understood in standard sense
\beq
A_{[µ_1µ_2...µ_p]}=\frac{1}{p!} \sum_{\sigma(µ_1µ_2...µ_p)} {\rm sgn}\sigma
A_{\sigma(µ_1)\sigma(µ_2)...\sigma(µ_p)}
\eeq
where summation is over all permutations of indices $µ_1µ_2...µ_p$ and the symbol
${\rm sgn}\sigma$ is the sign of given permutation.

The free action for such a field
in flat space-time
is described by the functional \cite{BKR}, \cite{Zin-2009}, \cite{LZ}
\footnote{Note that in Ref. \cite{BKR}, for the first time,  an action for
antisymmetric tensor-spinor fields has been constructed as well in $AdS$ space of arbitrary
dimensions.}
\beq
\label{Spex}
S_0[\psi]=-(-1)^{\frac{p(p-1)}{2}}\int d^n x \overline{\psi}_{µ_1µ_2...µ_p}
\Gamma^{µ_1µ_2...µ_p\nu \nu_1\nu_2...\nu_p}\pa_{\nu}\psi_{\nu_1\nu_2...\nu_p},
\eeq
where $\overline{\psi}=\psi^{\dag}\gamma^0$ and the notation
\beq
\Gamma^{µ_1µ_2...µ_p\nu \nu_1\nu_2...\nu_p}=\gamma^{[\mu_1}\gamma^{\mu_2}\cdots
\gamma^{\mu_p}\gamma^{\nu}\gamma^{\nu_1}\gamma^{\nu_2}\cdots \gamma^{\nu_p]}
\eeq
is used. $\gamma$-matrices satisfy the standard relations
\beq
\gamma^{\mu}\gamma^{\nu}+\gamma^{\nu}\gamma^{\mu}=2g^{\mu\nu}.
\eeq
The action (\ref{Spex}) can be considered as
a direct generalization of the Rarita-Schwinger action for a fermionic one-form
$\psi^a_{\mu}$
\beq
\label{SexRS}
S_0[\psi]=-\int d^n x \overline{\psi}_{µ}
\Gamma^{µ \nu \sigma}\pa_{\nu}\psi_{\sigma},
\eeq
which is invariant under the gauge transformations
\beq
\delta\psi^a_{\mu}=\pa_{\mu}\Lambda^a.
\eeq
The gauge generators
\beq
R^a_{\mu b}=\pa_{\mu}\delta^a_{\;\!b} ,\quad
\delta\psi^a_{\mu}=R^a_{\mu b}\Lambda^b
\eeq
do not depend on fields $\psi^a_{\mu}$, and
this simple model belongs to the class of gauge theories with
Abelian irreducible gauge algebra.
The free theory of fermionic 2-form tensor-spinor fields, $\psi^a_{\mu\nu}$, presents
a model of first-stage reducible gauge theory with action
\beq
\label{Spex2}
S_0[\psi]=\int d^n x \overline{\psi}_{µ\nu}
\Gamma^{µ\nu\rho\sigma\delta }\pa_{\rho}\psi_{\sigma\delta},
\eeq
being invariant under the following gauge transformations
\beq
\delta\psi^a_{\mu\nu}=2\pa_{[\mu}\Lambda^a_{\nu]},\qquad
\delta\Lambda^a_{\mu}=\pa_{\mu}\Lambda^a.
\eeq
The gauge generators
\beq
R^{a\sigma}_{\mu\nu b}= 2\pa_{[\mu}\delta^{\sigma}_{\;\!\nu]}\delta^a_{\;\!b},\qquad
\delta\psi^a_{\mu\nu}=R^{a\sigma}_{\mu\nu b}\Lambda^b_{\sigma},
\eeq
have the zero-eigenvalue eigenvectors
\beq
Z^a_{\mu b}=\pa_{\mu}\delta^a_{\;\!b},\qquad
R^{a\sigma}_{\mu\nu b}Z^b_{\sigma c}=0.
\eeq

It is clear that the action (\ref{Spex}) is disappeared if the dimension of space-time
satisfies the conditions $n\leq 2p$.
The action is invariant under  reducible gauge transformations. They are
\beq
\nonumber
&&
\delta\psi^a_{µ_1µ_2...µ_p}=p\pa_{[\mu_1}\Lambda^{(p-1)a}_{\quad\mu_2\cdots \mu_p]},\quad
\delta\Lambda^{(p-1)a}_{\quad\mu_2\cdots \mu_p}=
(p-1)\pa_{[\mu_2}\Lambda^{(p-2)a}_{\quad\mu_3\cdots \mu_p]},\\
\label{ginvpf}
&&\delta\Lambda^{(p-2)a}_{\quad\mu_3\cdots \mu_p}=
(p-2)\pa_{[\mu_3}\Lambda^{(p-3)a}_{\quad\mu_4\cdots \mu_p]}, \quad. . . ,
\quad
\delta\Lambda^{(1)a}_{\mu}=\pa_{\mu}\Lambda^{(0)a}.
\eeq
where $\Lambda^{(k)a}_{\quad\mu_1\cdots\mu_k}$, $k=0,1,...,p-1$,
is a rank-k antisymmetric tensor-spinor.
From (\ref{ginvpf}), it follows the identification for gauge generators
\beq
R^{a \nu_2\cdots \nu_p}_{\mu_1\mu_2\cdots \mu_p b}=
p\pa_{[\mu_1}\delta^{\nu_2}_{\mu_2}\cdots \delta^{\nu_p}_{\mu_p]}\delta^{a}_{\;\!b},
\eeq
and for the set of zero-eigenvalue eigenvectors
\beq
Z^{a \nu_2\cdots \nu_{s-1}}_{\mu_1\mu_2\cdots \mu_{s-1} b}=
(s-1)\pa_{[\mu_1}\delta^{\nu_2}_{\mu_2}\cdots \delta^{\nu_{s-1}}_{\mu_{s-1}]}
\delta^{a}_{\;\!b}, \quad s=2,3,...,p,
\eeq
in such a way that the relations (\ref{GTR1}), (\ref{GTR2}) in general setting
are reading now as
\beq
R^{a \sigma_2\cdots \sigma_{p}}_{\mu_1\mu_2\cdots \mu_p c}
Z^{c \nu_2\cdots \nu_{p-1}}_{\sigma_2\cdots \sigma_{p} b}=0,\qquad
Z^{a \sigma_2\cdots \sigma_{s-1}}_{\mu_1\mu_2\cdots \mu_{s-1} c}
Z^{c \nu_2\cdots \nu_{s-2}}_{\sigma_2\cdots \sigma_{s-1} b}=0,\quad s=3,...,p .
\eeq
Therefore, we have the free $(p-1)$-stage reducible
gauge theory.

We are going to study consistent deformations of the action (\ref{Spex}) in a way
describing above.
To do this correctly, we give the table of "quantum" numbers of quantities entering
in presentation  of action (\ref{Spex}) and used later:
\begin{center}
\begin{tabular}[c]{|c|c|c|c|c|c|c|c|}
\hline
Quantity & $\psi,\overline{\psi}$
 & $d^nx$ & $\pa_{\nu}$ &$\Gamma$& $\Box $ & $\varphi $ & m\\
\hline
$\varepsilon$ & 1&  0& 0 & 0 & 0 & 0 & 0\\
\hline
{\rm gh} & 0& 0& 0 &0 & 0 & 0 & 0 \\
\hline
dim & (n-1)/2& -n& 1&0 & 2 & (n-2)/2& 1\\
\hline
$\varepsilon_f $ &  1,-1&  0& 0& 0 &0 &0 & 0  \\
\hline
\end{tabular}
\end{center}
where $"\varepsilon"$ describes the Grassmann parity, the symbol
 $"{\rm gh}"$ is used to denote the ghost number,
 $"{\rm dim}"$ means the canonical dimension and   $"\varepsilon_f"$ is the
fermionic number. Using the table of "quantum" numbers, it is easy to establish
the quantum numbers of any quantities met in this section.

Deformation of initial theory is described by the generating function
$h^a_{µ_1µ_2...µ_p}(\psi)$ having the same "quantum" numbers as $\psi^a_{µ_1µ_2...µ_p}$.
Due to $\varepsilon_f(\psi)=1$ the generating function $h_{µ_1µ_2...µ_p}(\psi)$
should be a polynomial containing in each their term even number, say  $2k$, of fields
$\psi$ and, therefore,
odd number $2k-1$ of fields $\overline{\psi}$.
Such structure of terms $(\overline{\psi})^{2k-1}(\psi)^{2k}$ leads automatically to
the relation $\varepsilon(h)=1$. Canonical dimension of product of fields is equal to
\beq
{\rm dim}((\overline{\psi})^{2k-1}(\psi)^{2k})=(4k-1)\frac{(n-1)}{2}.
\eeq
To arrive at the needed relation ${\rm dim}(h)=(n-1)/2$, we have to use the dimensional
quantities $\pa$ and $\Box =\pa^{\nu}\pa_{\nu}$ in the term under consideration.
If the term contains $l$ partial derivatives, then one needs to introduce the operator
$\Box$ in the negative power $(2k-1)(n-1)+l/2$. Moreover, the function $h_{µ_1µ_2...µ_p}$
should be an antisymmetric  tensor-spinor field. The simple example of generating function
$h_{µ_1µ_2...µ_p}(\psi)$ satisfying all listed requirements
and corresponding to the case $k=1$ and
$l=1$ reads
\beq
h_{µ_1µ_2...µ_p}(\psi)=\frac{1}{\Box^{\frac{n}{2}}}
\psi_{µ_1µ_2...µ_p}\overline{\psi}_{\nu_1\nu_2...\nu_p}
\Gamma^{\nu_1\nu_2...\nu_p\nu \rho_1\rho_2...\rho_p}
\pa_{\nu}\psi_{\rho_1\rho_2...\rho_p}, \qquad n>2p.
\eeq
In the case of the Rarita-Schwinger action (\ref{SexRS}) it means
\beq
h_{µ}(\psi)=\frac{1}{\Box^{\frac{n}{2}}}
\psi_{µ}\overline{\psi}_{\nu}
\Gamma^{\nu\sigma \rho}
\pa_{\sigma}\psi_{\rho},
\eeq
as well as for the fermionic 2-form tensor-spinor fields (\ref{Spex2}) the generating
function has the form
\beq
h_{µ\nu}(\psi)=\frac{1}{\Box^{\frac{n}{2}}}
\psi_{µ\nu}\overline{\psi}_{\alpha\beta}
\Gamma^{\alpha\beta\delta\sigma\rho}
\pa_{\delta}\psi_{\sigma\rho}.
\eeq
Minimal dimension of space-time in the Rarita-Schwinger model is equal to 3.
Therefore, we can conclude that the non-locality
of deformed action comes from the fourth order vertex due to presence of operator $1/\Box$
and from the sixth order vertex because of $(1/\Box)^2$.
As the dimension of space - time grows, so does the degree of the operator $1/\Box$
responsible for the non-locality
of the deformed action. Analogous statement
about the non-locality is valid for the deformed model of
fermionic 2-form tensor-spinor fields.
In general, any consistent gauge-invariant deformation of antisymmetric
tensor-spinor fields creates a non-local deformed action which has no some closed
local sector. Let us stress once again that appearance of the operator
$(1/\Box)$ in the generating function $h_{µ_1µ_2...µ_p}(\psi)$ is dictated
by the strong motivations, namely : 1) non-triviality of the deformed gauge action
requires to use a non-local generating functions because generating local functions
with higher derivatives are forbidden by dimension reasons, 2) generating functions must
be non-linear in fields to reproduce vertexes of interactions, 3) preservation
of the fermionic number restricts possible non-linearity in fields
of the generating functions which must contain odd orders of fields
in its Taylor expansion,  4) compensation for the growing positive dimension
of the terms in the generating function
containing fields can be achieved using the corresponding positive
powers of the operator $(1/\Box)$.
The situation differs from the case of Abelian vector field or
massless bosonic higher spin fields \cite{BL-1} when the non-locality of generating
functions due to the operator $(1/\Box)$ is responsible for existence of
the local gauge sectors
of deformed actions which are gauge invariant under local pieces of the deformed
gauge generators.

\section{Interactions of fermionic 2-form fields and scalar field}
\noindent
Now we are going to demonstrate how the introduction of new degree of freedom in the form of
a scalar field may change the conclusion given in the previous section about non-local
nature of interactions of fermionic p-form fields.

We start with the free action of
fermionic 2-form fields, $\psi^a_{µ\nu}$, and a real massive scalar field, $\varphi$,
\beq
S_0[\psi,\varphi]&=&\int d^n x \overline{\psi}_{µ_1µ_2}
\Gamma^{µ_1µ_2\nu \nu_1\nu_2}\pa_{\nu}\psi_{\nu_1\nu_2}
+\frac{1}{2}\int d^nx\big(\pa^{\mu}\varphi\pa_{\mu}\varphi -m^2\varphi^2\big), \quad
n> 4 \;\!.
\label{iafs}
\eeq
The action is invariant under the gauge transformations
\beq
\delta\psi_{\mu\nu}=2\pa_{[\mu}\Lambda_{\nu ]},\quad \delta\varphi=0,
\eeq
and belongs to gauge fields of  first-stage reducibility with zero-eigenvalue eigenvectors
which do not depend on fields and, therefore, do not transform under deformations described
in section 3.
In this case, the identification with general notations begins with fields
$A^i=(\psi_{\mu\nu},\varphi)$ and generating functions of
anticanonical transformations
$h^i(A)=(h_{\mu\nu}(\psi,\varphi), h(\psi,\varphi))$.

The deformation of initial classical system (\ref{iafs}) is determined by arbitrary  choice of
generating functions with only restrictions concerning "quantum numbers" and
transformation rules
which should coincide with properties of corresponding fields so that
\beq
&&{\rm dim}(h_{\mu\nu})=\frac{n-1}{2},
\quad {\rm gh}(h_{\mu\nu})=0,\quad \varepsilon(h_{\mu\nu})=1,\quad
\varepsilon_f(h_{\mu\nu})=1,\\
&&{\rm dim}(h)=\frac{n-2}{2},\quad {\rm gh}(h)=0,\quad \varepsilon(h)=0,\quad
\varepsilon_f(h)=0 ,
\eeq
and $h$ must be a real scalar function while $h_{\mu\nu}$ must be a fermionic 2-form fields.

It is not difficult to propose the generating functions of anticanonical transformations
which will be responsible to generate cubic vertexes
in lower order of the deformation procedure,
\beq
h_{\mu\nu}(\psi,\varphi)=
g(m)^{\frac{4-n}{2}}\frac{1}{\Box}\pa_{\alpha}\big(\gamma^{\alpha}\psi_{\mu\nu}\varphi\big),
\qquad
h(\psi,\varphi)=0.
\eeq
The deformed action, $\widetilde{S}_0[\psi,\varphi]$, can be presented in the form
\beq
\label{defac}
\widetilde{S}_0[\psi,\varphi]&=&S_0[\psi,\varphi] +2g(m)^{\frac{4-n}{2}}\int d^nx
\overline{\psi}_{µ\nu}
\Gamma^{µ\nu\alpha\beta}\psi_{\alpha\beta}\varphi +\\
\nonumber
&&+{(non-local\;\; interaction \;\;terms)}.
\eeq
In deriving (\ref{defac}), the relations
\beq
&&\Gamma^{µ_1µ_2\nu \nu_1\nu_2}\gamma^{\mu}+
\Gamma^{µ_1µ_2\mu \nu_1\nu_2}\gamma^{\nu}=
2\;\!\Gamma^{µ_1µ_2\nu_1\nu_2}g^{\mu\nu}+\\
\nonumber
&&\qquad+
{(terms\;\; responsible \;\;for \;\;non-local \;\;contributions)},
\eeq
were used. The action
\beq
S_1[\psi,\varphi]=S_0[\psi,\varphi] + S_{int}[\psi,\varphi], \qquad
S_{int}[\psi,\varphi]= 2g(m)^{\frac{4-n}{2}}\int d^4x
\overline{\psi}_{µ\nu}
\Gamma^{µ\nu\alpha\beta}\psi_{\alpha\beta}\varphi
\eeq
describes the local sector of the deformed initial system. In its turn, the deformed gauge
transformations of fields $\psi_{\mu\nu}$ read
\beq
\widetilde{\delta}\psi_{\mu\nu}=\delta\psi_{\mu\nu}+\delta_1\psi_{\mu\nu}+O(g^2),
\eeq
where
\beq
\delta_1\psi_{\mu\nu}=-g(m)^{\frac{4-n}{2}}\frac{1}{\Box}\gamma^{\sigma}\pa_{\sigma}
(\varphi\delta\psi_{\mu\nu}).
\eeq
Let us consider the variation of action $S_1[\psi,\varphi]$ under the gauge transformations
$\overline{\delta}\psi_{\mu\nu}=\delta\psi_{\mu\nu}+\delta_1\psi_{\mu\nu}$,
\beq
\overline{\delta}S_1[\psi,\varphi]=\delta_1S_0[\psi,\varphi]+
\delta S_{int}[\psi,\varphi]+O(g^2).
\eeq
We have
\beq
\label{ld1}
&&\delta_1S_0[\psi,\varphi]=-g(m)^{\frac{4-n}{2}}\int d^4x\overline{\psi}_{µ\nu}
\Gamma^{µ\nu\alpha\beta}\delta\psi_{\alpha\beta}\varphi + {(non-local\;\; terms)},\\
\label{ld2}
&&\delta S_{int}[\psi,\varphi]=g(m)^{\frac{4-n}{2}}\int d^4x\overline{\psi}_{µ\nu}
\Gamma^{µ\nu\alpha\beta}\delta\psi_{\alpha\beta}\varphi.
\eeq
From Eqs. (\ref{ld1}), (\ref{ld2}), it follows that the local action $S_1[\psi,\varphi]$
describes interactions between fermionic 2-form fields and real massive scalar field
and is invariant
in the first order of deformation parameter under the gauge transformations
$\overline{\delta}\psi_{\mu\nu}$ up to non-local terms.
We see that construction of a local gauge theory of interacting antisymmetric spin-tensor
fermionic fields meets with  certain difficulties even if one introduces new degrees
of freedom in the form of massive scalar field: although,  in contrast with case which
was studied in the previous section,
the deformed action contains a local part
but it is not invariant under  local gauge symmetries.
We can conclude that up to now, description of interactions of fermionic p-form fields
in terms of a local gauge theory remains open problem.

\section{Conclusion}
\noindent
In the present paper,  we have extended the new approach proposed in \cite{BL-1},
\cite{BL-2} for gauge theories with closed/open algebras
to the procedure of gauge-invariant deformations of classical
reducible gauge theories. The deformation procedure of an initial
theory with gauge freedom
can be embedded into solutions to the classical master equation
of the BV formalism \cite{BH}, \cite{H}. Instead of using the cohomological approach
to find solutions to the classical master equation in the form of Taylor expansion
with respect to a deformation parameter  \cite{BH}, \cite{H},
it was proposed to take into account the invariance of the classical master equation
under anticanonical transformations. It allows to convert a given initial gauge theory
presented in the BV formalism through an action satisfying the classical master equation and
the boundary condition involving the initial gauge action into solutions
to the classical master equation containing full information about  deformed gauge-invariant
theory \cite{BL-1}, \cite{BL-2}.

Using analysis of anticanonical transformations in process of gauge-invariant deformation
of solutions to the classical master equation in the minimal antisymplectic space
\cite{BL-1}, \cite{BL-2}, we have made use of a single generating function
$h(A)$ depending on fields of initial theory $A$ only. It means that non-trivial part of the
generating functional $Y$ has the form $A^*h(A)$. In general,
the generating functional $Y$ of anticanonical transformations may contain terms of higher
order in antifields $(A^*)^m(C^*)^n(C^*_1)^kH_{m,n,k}(A)$.
Using $H_{m,n,k}(A)$ when at least the index $m>1$ does not change
the structure of deformation in the initial configuration space
but leads only to redefinition of
structure functions in the deformed gauge algebra. It is well-known fact
that the structure functions of any gauge algebra are
not define uniquely \cite{BV}, \cite{BV1}.
In fact, this arbitrariness has been fixed by special type of anticanonical
transformations in our method.
The deformation of initial action has the form of
replacement in the initial action the gauge field $A$ by
the field $A + h(A)$. In particular, it means that the generating function
$h(A)$ should be a non-local one or/and should contain higher derivatives because
otherwise one meets with trivial deformation when the deformed theory is classically
equivalent with the initial gauge system. In general, the deformed gauge theory is non-local
but sometimes it may happen that there exists a local gauge-invariant  sector
as a part of full theory. At the present, we have two important examples of such situation,
namely, the deformation of free Abelian gauge theory allows
to reproduce the Yang-Mills theory as well as the suitable non-local deformation
of free theory of massless bosonic higher spin fields \cite{Fronsdal-1}
leads to generation of
all local cubic vertexes known in the literature \cite{metsaev-1}, \cite{manvelyan},
\cite{metsaev}, \cite{zinoviev}(for discussions of the non-locality of higher order
vertexes, see \cite{Taronna}, \cite{Tsulaia}, \cite{Taronna-1}, \cite{Tseytlin},
\cite{Ponomarev}, \cite{Vasiliev}).

The deformation of gauge generators  is described
by the same function $h(A)$ in the form of shift $A\rightarrow A+h(A)$ of the argument
of initial gauge generators followed by
rotation defining by the inverse matrix to the
$M^i_{\;j}(A)=\delta^i_{\;j}+h^i_{\;\!,j}(A)$.
The deformation of zero-eigenvalue eigenvectors is described as the shift
$A\rightarrow A+h(A)$ in their arguments.
We have calculated some lower relations in the deformed reducible algebra
with deformed structure coefficients in the case of first-stage reducibility.
Generalization to arbitrary $L$-stage reducible gauge algebra looks like
as a technical task.
We emphasize that the deformed gauge algebra belongs to the same class of
reducible algebras as for the initial gauge algebra.

We have studied the free fermionic $p$-form fields as an example of reducible
gauge theory subjected to suitable gauge-invariant deformation. We have proved that
consistent self-interactions of these fields are always described  by non-local vertexes.
Therefore, if we deal  with fermionic $p$-form fields only then there is no possibility
to construct a local gauge theory of interactions between these fields.
In principle, adding new degree of
freedom to a given dynamical system may change
some properties of deformed theories.
In fact, it was a reason for us to consider the model of free fermionic 2-form fields
and a massive scalar field subjected to a non-local
gauge-invariant deformation leading to  existence  of cubic vertexes in the deformed action.
It was shown that  in the first order with respect of the deformation parameter the deformed
action contains a local part with cubic interactions of fields but, unfortunately,
it is not invariant under local gauge symmetries. So, construction of a local gauge theory,
containing interactions between completely antisymmetric spin-tensor fields, remains unsolved.

\section*{Acknowledgments}
\noindent
The author thanks I.L. Buchbinder   for useful discussions of different aspects of
higher spin field theories. The work is supported
by Ministry of Education of the Russian Federation, project
FEWF-2020-0003.


\begin {thebibliography}{99}
\addtolength{\itemsep}{-8pt}

\bibitem{BL-1}
I.L. Buchbinder, P.M. Lavrov,
\textit{On a gauge-invariant deformation of a classical gauge-invariant
theory}, JHEP 06 (2021) 854,  arXiv:2104.11930 [hep-th].


\bibitem{BL-2}
I.L. Buchbinder, P.M. Lavrov,
\textit{On classical and quantum deformations   of gauge theories},
Eur. Phys. J. {\bf C} 81 (2021) 856,
arXiv:2108.09968 [hep-th].

\bibitem{BV} I.A. Batalin, G.A. Vilkovisky, \textit{Gauge algebra and
quantization}, Phys. Lett. \textbf{B} 102 (1981) 27- 31.

\bibitem{BV1} I.A. Batalin, G.A. Vilkovisky, \textit{Quantization of gauge
theories with linearly dependent generators}, Phys. Rev. \textbf{D}
28 (1983) 2567-2582.

\bibitem{BV2} I.A. Batalin, G.A. Vilkovisky,
\textit{Closure of the gauge algebra,generalized Lie algebra
equations and Feynman rules}, Nucl. Phys. \textbf{\bf B} 234 (1984) 106.

\bibitem{VLT}
B.L. Voronov, P.M. Lavrov, I.V. Tyutin,
\textit{ Canonical transformations and gauge dependence
in general gauge theories},
Sov. J. Nucl. Phys. 36 (1982) 292-.

\bibitem{BLT-15}
I.A. Batalin, P.M. Lavrov, I.V.Tyutin, \textit {Finite anticanonical
transformations in field-antifield formalism}, Eur. Phys. J. {\bf C}
75 (2015) 270, {arXiv:1501.07334 [hep-th]}.

\bibitem{BL-16}
I.A. Batalin, P.M. Lavrov, \textit {Closed description of
arbitrariness in resolving quantum master equation}, Phys. Lett.
{\bf B} 758 (2016) 54-58, {arXiv:1604.01888 [hep-th]}.

\bibitem{BLT-21}
I.A. Batalin, P.M. Lavrov, I.V.Tyutin, \textit {Anticanonical
transformations and Grand Jacobian},
Russ. Phys. J. 64 (2021) 688-694, {arXiv:2011.06429 [hep-th]}.

\bibitem{AT}
 A. Andrasi, J.C. Taylor,
\textit {Generating functions for anti-canonical transformations
in the Zinn-Justin and Batalin and Vilkoviski formalisms},
{arXiv:2201.02106 [hep-th]}.

\bibitem{BH}
G. Barnich, M. Henneaux, \textit{Consistent coupling between fields
with gauge freedom and deformation of master equation}, Phys. Lett.
{\bf B} 311 (1993) 123-129, {arXiv:hep-th/9304057}.

\bibitem{H}
M. Henneaux, \textit{Consistent interactions between gauge fields:
The cohomological approach}, Contemp. Math. {\bf 219} (1998) 93-110,
{arXiv:hep-th/9712226}.

\bibitem{D}
A. Danehkar, \textit{On the cohomological derivation of Yang-Mills
theory in the antifield formalism}, JHEP,
Grav.Cosmol. 03 (2017) 368-387, {arXiv:0707.4025 [physics.gen-ph]}.

\bibitem{SS}
M. Sakaguchi, H. Suzuki,
\textit{On the interacting higher spin bosonic gauge fields in BRST-antifield
formalism}, Prog. Theor. Exp. Phys. 2021 (2021) 4, 043B01,
{arXiv:2011.02689 [hep-th]}.


\bibitem{DeWitt}
B.S. DeWitt, \textit{Dynamical theory of groups and fields},
(Gordon and Breach, 1965).

\bibitem{brs1}
C. Becchi, A. Rouet, R. Stora, \textit{The abelian Higgs Kibble
Model, unitarity of the $S$-operator}, Phys. Lett.  {\bf B} 52
(1974) 344- 346.

\bibitem{t}
I.V. Tyutin, \textit{Gauge invariance in field theory and
statistical physics in operator formalism}, Lebedev Institute
preprint  No. 39 (1975), arXiv:0812.0580 [hep-th].

\bibitem{BKR}
I.L. Buchbinder, V.A. Krykhtin, L.L. Ryskina,
\textit{Lagrangian formulation of massive fermionic totally antisymmetric
tensor field theory in $AdS_d$ space},
Nucl. Phys.  {\bf B819} (2009) 453-477, arXiv:0902.1471 [hep-th].

\bibitem{Zin-2009}
Yu.M. Zinoviev, \textit{Note on antisymmetric spin-tensors},
JHEP 04 (2009) 035, {arXiv:0903.0262 [hep-th]}.

\bibitem{LZ}
V. Lekeu, Yi Zhang,
\textit{On the quantisation and anomalies of antisymmetric tensor-spinors},
JHEP, {\bf 11} (2021) 078,
{arXiv:2109.03963 [hep-th]}.


\bibitem{Fronsdal-1}
C. Fronsdal, {\it Massless field with integer spin}, Phys. Rev. {\bf
D18} (1978) 3624.

\bibitem{metsaev-1}
R.R. Metsaev, \textit{Cubic interaction vertices of massive and
massless higher spin fields}, Nucl. Phys. {\bf B} 759 (2006)
141-201, {arXiv:hep-th/0512342}.

\bibitem{manvelyan}
R. Manvelyan, K. Mkrtchyan, W. Ruehl, \textit{A generating function
for the cubic interactions of higher spin fields},  Phys. Lett. {\bf
B} 696 (2011) 410-415, {arXiv:1009.1054 [hep-th]}.

\bibitem{metsaev}
R.R. Metsaev, \textit{BRST-BV approach to cubic interaction for
massive and massless higher spin fields}, Phys. Lett. {\bf B} 720
(2013) 237-243, {arXiv:1205.3131 [hep-th]}.

\bibitem{zinoviev}
M. V. Khabarov, Yu. M. Zinoviev, \textit{Massless higher spin cubic
vertices in flat four dimensional space},  JHEP {\bf 08} (2020) 112,
{arXiv:2005.09851 [hep-th]}.

\bibitem{Taronna}
M. Taronna, \textit{Higher-Spin Interactions: four-point functions
and beyond}, JHEP {\bf04} (2012) 029, {arXiv:1107.5843 [hep-th]}.

\bibitem{Tsulaia}
P. Dempster, M. Tsulaia, \textit{On the Structure of Quartic
Vertices for Massless Higher Spin Fields on Minkowski Background},
Nucl. Phys. {\bf B} 865 (2012) 353-375, {arXiv:1203.5597 [hep-th]}.

\bibitem{Taronna-1}
M. Taronna, \textit{On the Non-Local Obstruction to Interacting
Higher Spins in Flat Space}, JHEP {\bf 05} (2017) 026, {
arXiv:1701.05772 [hep-th]}.

\bibitem{Tseytlin}
R. Roiban, A.A. Tseytlin, \textit{On four-point interactions in
massless higher spin theory in flat space}, JHEP {\bf 04} (2017) 139,
{arXiv:1701.05773 [hep-th]}.

\bibitem{Ponomarev}
D. Ponomarev, \textit{A Note on (Non)-Locality in Holographic Higher Spin Theories},
Universe {\bf 4} (2018) 2, {arXiv:1710.00403 [hep-th]}.

\bibitem{Vasiliev}
O.A. Gelfond, M.A. Vasiliev, \textit{Spin-Locality of Higher-Spin
Theories and Star-Product Functional Classes},
JHEP {\bf 03} (2020) 002, {arXiv:1910.00487
[hep-th]}.



\end{thebibliography}
\end{document}